\documentclass{article}
\usepackage{spconf,amsmath,graphicx}
\usepackage{booktabs}
\usepackage{multirow}
\usepackage{bm}
\usepackage{booktabs}
\usepackage{hyperref}
\usepackage{threeparttable}
\usepackage[ruled]{algorithm2e}

\title{Improving Cross-lingual Speech Synthesis with Triplet Training Scheme}
\name{Jianhao Ye, Hongbin Zhou, Zhiba Su, Wendi He, Kaimeng Ren, Lin Li, Heng Lu}
\address{Ximalaya Inc., China\\$\{jianhao.ye, hongbin.zhou, zhiba.su, cloris.he, irving.ren, lin.li, bear.lu\}$@ximalaya.com}

\begin{document}



%
%
%

%
\maketitle
\begin{abstract}

Recent advances in cross-lingual text-to-speech (TTS) made it possible to synthesize speech in a language foreign to a monolingual speaker. However, there is still a large gap between the pronunciation of generated cross-lingual speech and that of native speakers in terms of naturalness and intelligibility. In this paper, a triplet training scheme is proposed to enhance the cross-lingual pronunciation by allowing previously unseen content and speaker combinations to be seen during training. Proposed method introduces an extra fine-tune stage with triplet loss during training, which efficiently draws the pronunciation of the synthesized foreign speech closer to those from the native anchor speaker, while preserving the non-native speaker's timbre. Experiments are conducted based on a state-of-the-art baseline cross-lingual TTS system and its enhanced variants. All the objective and subjective evaluations show the proposed method brings significant improvement in both intelligibility and naturalness of the synthesized cross-lingual speech.

\end{abstract}

\vspace{-5pt}


\vspace{-5pt}

\section{Introduction}

\vspace{-5pt}

\label{sec:intro}


Building a cross-lingual TTS system is a task that requires the system to synthesize speech of a language foreign to a target speaker \cite{xin_disentangled_2021}. It is straightforward to build a multi-lingual TTS system with a multi-lingual corpus using end-to-end models \cite{shen_natural_2018, yu_durian_2019, ren_fastspeech_2021}, yet such corpus of a same speaker is often difficult to collect. Therefore, much efforts have been made on building a cross-lingual TTS system with only monolingual data. 

The majority of previous studies focuses on building cross-lingual TTS system by mixing monolingual corpora of different languages, while disentangling speaker representations and language or phonetic representations. \cite{zhang_learning_2019} uses adversarial training to remove speaker identity from phonetic representations. \cite{xin_disentangled_2021} employs mutual information minimization and domain adaptation to disentangle the learned language and speaker embedding, which improves naturalness and speaker similarity of the synthesized cross-lingual speech. Beside explicit disentanglement, others use unified phonetic representations to share pronunciation across languages, which helps to implicitly disentangle language and speaker. \cite{he_multilingual_2021, li_bytes_2018} and \cite{zhao_towards_2020, cao_code-switched_2020} use Unicode bytes and Phonetic Posterior-Grams (PPGs) respectively as common phonetic set to build cross-lingual TTS systems and all get improvements compared to their baselines. Moreover, \cite{hemati_using_2020, zhan_improve_2021} convert all the graphemes of different languages into a same International Phonetic Alphabet (IPA) set to facilitate cross-lingual modeling and the experiments in \cite{zhan_improve_2021} show the privilege of IPA over language-dependent phonemes.




In most of the above works, the cross-lingual combination of a target speaker and a foreign language in inference is still unseen during training, leaving a gap between inference and training. Similar gap also exists in the task of TTS style transfer. \cite{whitehill_multi-reference_2020} narrows the gap by introducing multiple reference encoders that encode both the synthesized and ground-truth (GT) acoustic features and allowing gradients flow through the synthesized ones to cover unseen cases in training. In \cite{xin_cross-lingual_2021}, a consistency loss over speaker identity is applied by computing distance of speaker embeddings of the synthesized and the GT mel-spectrogram, which helps to keep target speaker identity after the cross-lingual transfer. 

Inspired by \cite{xin_cross-lingual_2021} and \cite{whitehill_multi-reference_2020}, we employ triplet loss \cite{schroff_facenet_2015} and propose a triplet training scheme to cover unseen cases. A triplet is composed of an anchor, a positive and a negative sample. The triplet loss is a pairwise loss that minimizes the distance between the representation of anchor sample and that of positive sample while maximizing the distance between anchor and the negative one \cite{schroff_facenet_2015}. In the proposed method, triplet loss encourages the pronunciation of the synthesized foreign speech to be close to that of the native anchor speaker, while maintaining the speaker identity of its own. The proposed triplet training scheme has two stages. The first stage is a regular from-scratch training but with auxiliary Content Predictor (CP) and Speaker Predictor (SP) which are used to extract representations for triplets. The second stage fine tunes the checkpoint from first stage by incorporating a triplet loss with triplets made from both synthesized and GT mel-spectrogram.

\begin{figure*}[ht]
\centering
\includegraphics[width=17cm]{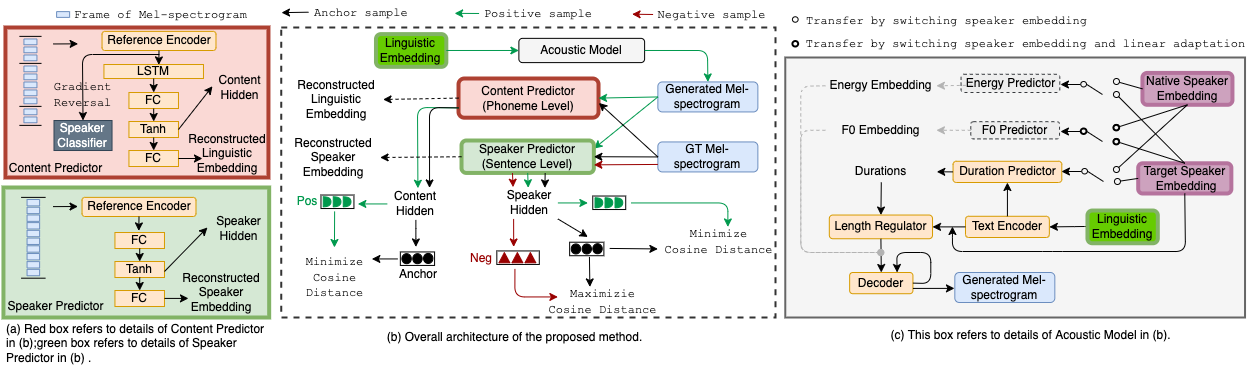}
\vspace{-10pt}
\caption{The overall architecture of cross-lingual text-to-speech system with triplet loss.} \label{fig:1}
\vspace{-10pt}
\end{figure*}

To perform evaluations, we train DurIAN \cite{yu_durian_2019} both with and without triplet training scheme, using monolingual corpora of English (EN) and Mandarin (CN). For further investigation, we also apply the proposed method to an enhanced version of DurIAN capable of transferring phoneme-level prosody features: f0, energy and duration. The results of objective and subjective evaluations show that the proposed method brings improvements in both intelligibility and naturalness to cross-lingual speech, while maintaining speaker identity. Some of the evaluated audio samples can be found in \url{https://jianhaoye2.github.io/icassp2022triplet.github.io/}.

\vspace{-5pt}

\section{Triplet Training Scheme}

\vspace{-5pt}


\label{sec:triplet_training}

\subsection{Construction of Content Triplet (CT) and Speaker Triplet (ST) and Triplet Loss}
\label{ssec:triplet_content}

\vspace{-5pt}

The proposed triplet training scheme consists of two stages with a triplet loss used in stage II. To compose the triplet loss, CT and ST need to be constructed with the help of CP and SP. In the definition of triplet loss \cite{schroff_facenet_2015}, there is an encoding function $f(.)$ which encodes data points into an embedding and a distance function $D(.)$ that measures the distance between a pair of encoded embeddings. In our work, cosine distance is selected as $D(.)$, while CP and SP work as encoding function as shown in Fig. \ref{fig:1} (a). CP gets input from phoneme-level mel-spectrogram and reconstructs linguistic embeddings of each phoneme. Specifically, first a mel-spectrogram $\bm{M}$ is split into segments $[\bm{m_1}, ..., \bm{m_{T_{ph}}}]$ with given phoneme duration, where $T_{ph}$ is the number of phonemes in current sequence. Second, each segment is encoded by a reference encoder and a speaker adversarial training task is used to remove speaker-related information from the encoding and the same speaker loss with gradient reversal as in \cite{zhang_learning_2019} is used. Then, encoded segments are further processed and the linguistic embedding of each phoneme $\bm{e^{C}_{t}}$ is finally reconstructed. The hidden $z^{C}_{t}$ before the last fully-connected (FC) layer is picked as encoded embedding of CT, which presumably contains pronunciation-related information since pronunciation is highly related to content. The encoding function of CT is summarized by $\bm{z^{C}_{t}} = f^{C}(\bm{m_{t}})$. Unlike CP, SP reconstructs speaker embedding $\bm{e^{S}}$ and its reference encoder, sharing the same structure as in CP, encodes the whole sequence of mel-spectrogram into a single representation. Similarly, the encoding  $\bm{z^{S}}$ before last FC layer is selected as encoded embedding for ST, whose encoding function is denoted by $\bm{z^{S}} = f^{S}(\bm{M})$.









Beside the encoding functions, pairs of CT and ST, designed to cover inter-language (inter-lan) or unseen cases in training, need to be constructed. Since CT aims to make phoneme-level pronunciation of the synthesized cross-lingual speech closer to that of the selected anchor native speaker, only positive samples -- speech with the same content as anchor sample are considered. CT is defined as $t^{C}(\bm{M^{an}}, \bm{\hat{M}^{pos}}, None)$, where $\bm{\hat{M}^{pos}} = [{\bm{\hat{m}_{1}^{pos}}, ..., \bm{\hat{m}^{pos}_{T_{ph}}}}]$. In a CT, the positive sample $\bm{\hat{M}^{pos}}$ refers to synthesized mel-spectrogram of a sentence foreign to the target speaker and with the same content with the anchor sample. The anchor sample $\bm{M^{an}}$ is the GT version of the same sentence from selected anchor speaker, and the anchor speaker is native to the target language. Similarly, ST is defined as $t^{S}(\bm{M^{an}_{S}}, \bm{\hat{M}^{pos}}, \bm{M^{neg}}_{S})$ to consider both similarity and dissimilarity to better maintain target speaker's timbre. In ST, $\bm{M^{an}_{S}}, \bm{\hat{M}^{pos}}, \bm{M^{neg}_{S}}$ are defined as ``anchor sample", ``positive sample", and ``negative sample" respectively in terms of speaker identity. Our implementation is illustrated in Alg. \ref{alg:triplets}.







%


\vspace{-15pt}

\begin{algorithm}
    \begin{small}   
    \caption{Triplets construction in a batch}
    \label{alg:triplets}
    \textbf{Input}: A batch of data $B$ composed of linguistic features $\bm{t}$, GT mel-spectrogram $\bm{M}$, speaker ID $s$, language $l$. An empty list $triplets\_list$.     
    
    \end{small}    
    
\begin{small}    
    \For{($\bm{t_i}, \bm{M_i}, s_{i}, l_i) \in B$} {
        \If{$s_{i}$ is the anchor speaker of $l_i$} {
            $s^{pos} \leftarrow$ Randomly pick a $s$ from $B$ (whose language != $l_i$)
            
            $\bm{\hat{M}^{pos}} \leftarrow Acoustic\_Model(\bm{t_i}, s^{pos})$ 
            
            $\bm{M^{an}_S}  \leftarrow$ Randomly pick a $\bm{M}$ from $B$ (whose speaker ID = $ s^{pos}$)
            
            $\bm{M^{neg}_S}  \leftarrow$ Randomly pick a $\bm{M}$ from $B$ (whose speaker ID != $s^{pos}$)
            
            $triplets\_list.append(t^{C}(\bm{M_i}, \bm{\hat{M}^{pos}}, None), $
            
            $t^{S}(\bm{M^{an}_S}, \bm{\hat{M}^{pos}}, \bm{M^{neg}_S}))$
        }
    }
\end{small}

    \textbf{Output}: $triplets\_list$

\end{algorithm}

\vspace{-15pt}

By applying encoding functions to constructed triplets as shown in Fig \ref{fig:1} (b), the triplet loss is finally defined in Eq. \ref{e_triplet_all}, where $\alpha$ and $\beta$ are weights that balance content and speaker parts and $t$ denotes phoneme index in a sequence. The margin parameter in \cite{schroff_facenet_2015} is removed since the main purpose of this task is not for classification.





\vspace{-10pt}

\subsection{Two Stage Triplet Training Scheme}

\vspace{-5pt}
The training of stage I is a regular from-scratch training like most acoustic models, but with these extra losses: adversarial training loss for CP, L2 reconstruction losses for CP and SP which are defined as $\pounds_{recon\_ling} = \sum_{t}^{T_{ph}} (\bm{e^{C}_{t}} - \bm{\hat{e}^{C}_{t}})^2$ and $\pounds_{recon\_spk} = (\bm{e^{S}} - {\bm{\hat{e}^{S}}})^2$. Considering DurIAN \cite{yu_durian_2019} as an example, the losses of which contain a reconstruction loss, a duration loss and a residual loss, defined respectively as $ \pounds_{recon} = \sum_{k}^{T_{f}} |\bm{m_k} - \bm{\hat{m_k}}|$, $\pounds_{dur} = \sum_{t}^{T_{ph}} (d_t - \hat{d_t})^2$, $ \pounds_{res} = \sum_{k}^{T_{f}} |\bm{m_k} - (\hat{\bm{m_k}} + \bm{r_k})|$ where the residual output $\bm{r_k}$ is the same as in \cite{yu_durian_2019} with $k$ as the frame index and $T_{f}$ as the number of total frames of a sequence. To sum up, the total loss is defined in Eq. \ref{e_step1_all} where the loss for speaker adversarial training is denoted by $\pounds_{adv}$ and $\lambda$ is the weight of speaker adversarial loss.

In stage II, fine-tune training is applied by loading the checkpoint from the first stage and fine-tune training with new losses. To keep the learned $f^{C}$ and $f^{S}$ unchanged, we keep the weights of speaker embedding, linguistic embedding, CP and SP fixed during this stage as shown in bold lines in Figure \ref{fig:1} (b)(c). Thus, $\pounds_{recon\_spk}$ and $\pounds_{recon\_ling}$ are no longer required. Meanwhile, the triplets are constructed and triplet loss is incorporated. Hence the final fine-tune loss as stated in Eq. \ref{e_train_ft}. Practically, We stop the fine-tune training when the triplet loss is converged, since other losses are already at convergence after stage I and remain converged or will converge very quickly in stage II.



\vspace{-10pt}

\begin{small}
\begin{equation}
\begin{aligned}
\pounds_{triplet} & = \alpha  * \max(0, \frac{ \sum_{t}^{T_ph} D(f^{C}(\bm{m^{an}_{t}}), f^{C}(\bm{\hat{m}^{pos}_{t}})) }{T_{ph}}) \\
& + \beta * \max(0, D(f^{S}(\bm{M^{an}}), f^{S}(\bm{\hat{M}^{pos}}))  \\
& - D(f^{S}(\bm{M^{an}}), f^{S}(\bm{M^{neg}})))
\end{aligned}
\label{e_triplet_all}
\end{equation}
\end{small}

\vspace{-25pt}

\begin{small}
\begin{align}
     \pounds_{fs} & = \pounds_{recon} + \pounds_{dur} + \pounds_{res} \nonumber \\
    & + \pounds_{recon\_ling} + \pounds_{recon\_spk} + \lambda * \pounds_{adv} \label{e_step1_all}    \\
        \pounds_{ft} & = \pounds_{recon} + \pounds_{dur} + \pounds_{res} +  \pounds_{triplet} \label{e_train_ft}
\end{align}

\end{small}








\section{experiments}









\vspace{-5pt}

\label{sec:exp}

\begin{table*}[ht]
\renewcommand\thetable{2}

\addtolength{\tabcolsep}{-5pt}

\centering
\caption{Naturalness and speaker similarity MOS with 95\% confidence interval.}

\begin{tabular}{cclcccccccclc}
\hline
Test utterance      & \multicolumn{7}{c}{EN utterance}                                                                                                                                &           & \multicolumn{4}{c}{CN utterance}                                                     \\ \cline{2-8} \cline{10-13} 
Speaker             & EN-SPK             &  & \multicolumn{2}{c}{\textit{\textbf{CN-SPK-M}}}               & \textbf{} & \multicolumn{2}{c}{\textit{\textbf{CN-SPK-F}}}               & \textbf{} & \multicolumn{2}{c}{\textit{\textbf{EN-SPK}}}                 &  & CN-SPK-F           \\ \cline{2-2} \cline{4-5} \cline{7-8} \cline{10-11} \cline{13-13} 
System              & Naturalness        &  & \textit{\textbf{Naturalness}} & \textit{\textbf{Similarity}} & \textbf{} & \textit{\textbf{Naturalness}} & \textit{\textbf{Similarity}} & \textbf{} & \textit{\textbf{Naturalness}} & \textit{\textbf{Similarity}} &  & Naturalness        \\ \cline{1-11} \cline{13-13} 
Base                & 3.28±0.12          &  & 2.98±0.11                     & 3.60±0.10                    & \textit{} & 3.15±0.14                     & 3.34±0.09                    & \textit{} & 2.73±0.12                     & 3.58±0.09                    &  & 4.19±0.08          \\
Base+Triplet        & \textbf{3.57±0.09} &  & \textbf{3.31±0.11}            & \textbf{3.72±0.14}           & \textit{} & \textbf{3.35±0.14}            & \textbf{3.78±0.09}           & \textit{} & \textbf{2.99±0.11}            & \textbf{3.73±0.07}           &  & \textbf{4.29±0.07} \\ \hline
Base\_FE\_DFE             & 3.98±0.08          &  & \textbf{3.33±0.14}            & 3.72±0.13                    & \textit{} & 3.39±0.11                     & 3.78±0.11                    & \textit{} & 3.20±0.09                     & 3.83±0.07                    &  & \textbf{4.33±0.06} \\
Base\_FE\_DFE+Triplet & \textbf{3.99±0.08} &  & 3.21±0.12                     & \textbf{3.80±0.07}           & \textit{} & \textbf{3.59±0.08}            & \textbf{3.82±0.07}           & \textit{} & \textbf{3.36±0.07}            & \textbf{3.93±0.07}           &  & 4.26±0.08          \\ \hline
Ground Truth        & 4.39±0.04          &  & /                             & /                            &           & /                             & /                            &           & /                             & /                            &  & 4.46±0.02          \\ \hline
\end{tabular}

\label{mos}

\vspace{-10pt}
\end{table*}


\vspace{-5pt}

\subsection{Systems}

\vspace{-5pt}


\label{ssec:Systems}








The overall architecture of models used for experiments is shown in Fig. \ref{fig:1} (c). 
Based on two baseline systems, the proposed triplet loss is implemented and compared. The first baseline system ``\textit{Base}" follows the DurIAN \cite{yu_durian_2019} structure, while its triplet version is system ``\textit{Base+Triplet}".  The enhanced baseline model ``\textit{Base\_FE}" is constructed by adding two extra prosody predictors of phoneme-level f0 and energy to \textit{Base}, which are demonstrated with dashed lines in Fig. \ref{fig:1} (c).  Those prosody predictors follow the same setup as in \cite{ren_fastspeech_2021} and quantize real values into trainable embeddings to condition decoder for synthesizing speech. And its triplet version is system ``\textit{Base\_FE+Triplet}".
Furthermore, since the triplet training scheme can be interpreted as a way to implicitly transfer phoneme-level prosody features, it is worthwhile to investigate the proposed method with transferring of explicit prosody features. Also, it is shown in \cite{hemati_using_2020} that replacing the phoneme alignment of target speaker with that of native speaker is beneficial to cross-lingual speech synthesis. Therefore, proposed methods are also tested with fine-grained explicit prosody transfer based on the \textit{Base\_FE} system. In system ``\textit{Base\_FE\_DFE}", the transfer of duration, f0, energy are implemented by replacing target speaker embedding with the native one in duration model, f0 predictor, energy predictor, only in the inference phase of cross-lingual cases, as shown in Fig. \ref{fig:1} (c). A linear adaptation is applied to the transferred f0 similar to Gaussian based mapping system in \cite{rallabandi_building_2017}.
While in system ``\textit{Base\_FE\_DFE+Triplet}", prosody transfer is used in both triplet training scheme and inference for cross-lingual cases, making the whole process more consistent. 


The training losses for \textit{Base} are described by Eq. \ref{e_step1_all} and Eq. \ref{e_train_ft} in Sec. \ref{sec:triplet_training}, while the losses for \textit{Base\_FE} include two extra L2 losses used to train f0 predictor and energy predictor.
 Therefore, the total losses of \textit{Base\_FE} for two stages of training become $\pounds_{fs\_f0\_energy} = \pounds_{fs} + \pounds_{f0} + \pounds_{energy}$ and $\pounds_{ft\_f0\_energy} = \pounds_{ft} + \pounds_{f0} + \pounds_{energy} + \pounds_{triplet}$.

\vspace{-10pt}

\subsection{Data Setup}

\vspace{-5pt}

\label{ssec:data}

Two public datasets are used in experiments: Data-baker Chinese Standard Mandarin Speech Corpus (CSMSC) \cite{csmsc}, a Mandarin dataset and LJSpeech dataset \cite{LJSpeech17}, an English speech dataset. The audio files in those datasets vary from 1 to 10 seconds and the total length is about 12 hours for  CSMSC and 24 hours for LJSpeech. As both public datasets are from female speakers, another Mandarin male dataset is included for more balanced evaluation, so that cross-gender cross-lingual speech synthesis is considered as well. The male dataset is recorded in studio and the total length of audio is almost equal to CSMSC. All audios are resampled to $24kHZ$. The speakers of the three datasets are denoted by EN-SPK, CN-SPK-F, CN-SPK-M for CSMSC, LJSPEECH and Mandarin male dataset.    

\vspace{-10pt}

\subsection{Experimental Setup}
\label{ssec:exp_sys}
\vspace{-5pt}

All the texts are converted into IPA symbols via espeak\footnote{\url{http://espeak.sourceforge.net/}}. 80-dimensional mel-spectrograms are extracted by using Hanning window with frame shift 10 ms and frame length 42.7 ms. Then, Kaldi toolkit \cite{povey_kaldi_nodate} is utilized to do forced-alignment. The extraction of GT f0 and energy follows \cite{ren_fastspeech_2021}. During the training, batch size is set to 16 for a single NVIDIA 2080Ti GPU and Adam Optimizer is used with learning rate equal to 0.0001. After cross-validations, the $\alpha$ and $\beta$ in Eq. \ref{e_triplet_all} are set as $1.0$ and $0.02$ respectively and the adversarial weight $\lambda$ is set to 0.025. In our setup, the average number of fine-tune steps is $50k$ for $\pounds_{triplet}$ to converge. Finally, a multi-speaker HiFi-GAN is trained to synthesize audio samples \cite{kong_hifi-gan_2020}. For CT construction, CN-SPK-F is selected as Mandarin anchor speaker, while EN-SPK is the English anchor speaker. 





\vspace{-10pt}

\subsection{Objective Evaluations}

\vspace{-5pt}

To evaluate the effectiveness of proposed triplet training scheme in  intelligibility, the speech recognition service of Azure\footnote{\url{https://azure.microsoft.com/}} is called to test Word Error Rate (WER) for generated cross-lingual speech. 100 sentences for each language are randomly selected from test set to produce test audios. Inter-lan tests of all speakers and intra-language (intra-lan) tests of CT anchor speakers are included in the experiments. 

Table \ref{spk3_1} shows the WER for all systems described in Sec.  \ref{ssec:Systems} and the results show systems with proposed training scheme (system ``\textit{Base+Triplet}", ``\textit{Base\_FE+Triplet}", ``\textit{Base\_FE\_DFE+Triplet}") consistently outperform all 3 baseline systems without triplet loss training (system ``\textit{Base}", ``\textit{Base\_FE}", ``\textit{Base\_FE\_DFE}") with a large margin. Moreover, proposed method performs comparatively stronger in inter-lan test cases than in intra-lan circumstances. That is reasonable since the CT in triplet loss is designed to work when there is a larger content gap in inter-lan cases than in intra-lan cases. Also, it is worth mentioning that although fine-grained explicit prosody transfer system \textit{Base\_FE\_DFE} does outperform the other two baselines, fine-tuning with proposed triplet loss still adds value, which makes training stage more consistent with inference stage. Moreover, \textit{Base\_FE\_DFE} fails to bring improvement on cross-gender cross-lingual test in terms of intelligibility but its triplet version \textit{Base\_FE\_DFE+Triplet} is proved to be effective on that test.

%

\begin{table}[h]
\renewcommand\thetable{1}

\vspace{-10pt}
\addtolength{\tabcolsep}{-5pt}
\centering
\caption{WER evaluation on intelligibility.}

\begin{tabular}{cccccccc}
\hline
Test utterance                                                 & \multicolumn{3}{c}{EN utterance}                                                                                                                                                                  &                      & \multicolumn{2}{c}{CN utterance}                                                                                          &                                    \\ \cline{2-4} \cline{6-7}
System                                                         & \begin{tabular}[c]{@{}c@{}}EN-\\ SPK\end{tabular} & \textit{\textbf{\begin{tabular}[c]{@{}c@{}}CN-\\ SPK-M\end{tabular}}} & \textit{\textbf{\begin{tabular}[c]{@{}c@{}}CN-\\ SPK-F\end{tabular}}} &                      & \textit{\textbf{\begin{tabular}[c]{@{}c@{}}EN-\\ SPK\end{tabular}}} & \begin{tabular}[c]{@{}c@{}}CN-\\ SPK-F\end{tabular} & AVG                                \\ \hline
\multicolumn{1}{l}{Ground Truth}                               & \multicolumn{1}{l}{\textbf{0.099}}                 & /                                                                     & /                                                                      & \multicolumn{1}{l}{} & /                                                                    & \multicolumn{1}{l}{\textbf{0.046}}                   & /                                  \\ \hline
Base                                                           & \multicolumn{1}{l}{0.122}                         & \multicolumn{1}{l}{0.261}                                             & \multicolumn{1}{l}{0.258}                                             & \multicolumn{1}{l}{} & \multicolumn{1}{l}{0.135}                                           & \multicolumn{1}{l}{\textbf{0.053}}                  & \multicolumn{1}{l}{0.166}          \\
Base+Triplet                                                   & \multicolumn{1}{l}{\textbf{0.119}}                & \multicolumn{1}{l}{\textbf{0.197}}                                    & \multicolumn{1}{l}{\textbf{0.212}}                                    & \multicolumn{1}{l}{} & \multicolumn{1}{l}{\textbf{0.097}}                                  & \multicolumn{1}{l}{0.056}                           & \multicolumn{1}{l}{\textbf{0.136}} \\ \hline
Base\_FE                                                        & \textbf{0.112}                                    & 0.243                                                                 & 0.261                                                                 &                      & 0.134                                                               & \textbf{0.049}                                      & 0.160                              \\
\begin{tabular}[c]{@{}c@{}}Base\_FE\\ +Triplet\end{tabular}     & 0.119                                             & \textbf{0.197}                                                        & \textbf{0.218}                                                        &                      & \textbf{0.107}                                                      & \textbf{0.049}                                      & \textbf{0.138}                     \\ \hline
Base\_FE\_DFE                                                    & /                                                 & 0.244                                                                 & 0.213                                                                 &                      & 0.086                                                               & /                                                   & 0.141                              \\
\begin{tabular}[c]{@{}c@{}}Base\_FE\_DFE\\ +Triplet\end{tabular} & 0.113                                             & \textbf{0.219}                                                        & \textbf{0.211}                                                        &                      & \textbf{0.075}                                                      & 0.052                                               & \textbf{0.134}                     \\ \hline
\end{tabular}

\label{spk3_1}
\begin{small}
\begin{tablenotes}
\item Inter-lan tests are emphasized with \textit{Italic} + \textbf{Bold}.
\end{tablenotes}
\end{small}
\vspace{-15pt}
\end{table}

\vspace{-10pt}

\subsection{Subjective Evaluations}

\vspace{-5pt}


Mean Opinion Scores (MOS) tests on both naturalness and cross-lingual speaker similarity are conducted for subjective evaluations following \cite{yu_durian_2019}. 50 sentences for each language are randomly picked from test set and 8 native raters per language participated in the tests. Beside \textit{Base} and \textit{Base+Triplet}, the enhanced system \textit{Base\_FE\_DFE} and its triplet version, which are better than system \textit{Base\_FE} and \textit{Base\_FE+Triplet} in terms of intelligibility, are also included in subjective evaluations.


The subjective MOS results are shown in Table \ref{mos}. The results are consistent with that of objective evaluations in two aspects. First, both \textit{Base+Triplet} and \textit{Base\_FE\_DFE+Triplet} gain improvements in naturalness and speaker similarity from the proposed triplet scheme, compared to their counterpart models without fine-tuning with triplet loss. By comparing \textit{Base} and \textit{Base+Triplet}, we also find that not only the inter-lan tests but also the intra-lan tests benefit from the proposed method. That could be related to the designed triplets alleviating the interference from the data of different languages when shared phonemes (IPA) are used in cross-lingual TTS. Second, further improvement is achieved by combining the fine-grained explicit prosody transfer with the proposed method, though the improvement of cross-gender cross-lingual test is relatively smaller.




\label{ssec:exp_obj}



\label{ssec:exp_subj}


\vspace{-5pt}

\section{conclusion}
\label{sec:conclude}

\vspace{-5pt}

This paper proposed a method that employs triplet loss to cover unseen combinations of target speaker and foreign text in cross-lingual TTS. The experimental results show the proposed method significantly enhances the intelligibility as well as naturalness of cross-lingual speech, while maintaining the target speaker identity. Moreover, we also find the proposed method can be combined with fine-grained explicit prosody transfer to further improve the performance of cross-lingual TTS, though the improvement to cross-gender cross-lingual cases is relatively small. Although the triplets in this work are constructed to narrow the gap between training and inference stages for cross-lingual purpose, the triplets can be redesigned to form a triplet loss for unseen cases in other tasks, which will be explored in the future.




\vfill\pagebreak

\bibliographystyle{IEEEbib}
\bibliography{main_ref}

\begin{thebibliography}{10}

\bibitem{xin_disentangled_2021}
Detai Xin, Tatsuya Komatsu, Shinnosuke Takamichi, and Hiroshi Saruwatari,
\newblock ``Disentangled speaker and language representations using mutual
  information minimization and domain adaptation for cross-lingual {TTS},''
\newblock in {\em {ICASSP} 2021 - 2021 {IEEE} International Conference on
  Acoustics, Speech and Signal Processing ({ICASSP})}. pp. 6608--6612, {IEEE}.

\bibitem{shen_natural_2018}
Jonathan Shen, Ruoming Pang, Ron~J Weiss, Mike Schuster, Navdeep Jaitly,
  Zongheng Yang, Zhifeng Chen, Yu~Zhang, Yuxuan Wang, Rj~Skerrv-Ryan, et~al.,
\newblock ``Natural tts synthesis by conditioning wavenet on mel spectrogram
  predictions,''
\newblock in {\em 2018 IEEE International Conference on Acoustics, Speech and
  Signal Processing (ICASSP)}. IEEE, 2018, pp. 4779--4783.

\bibitem{yu_durian_2019}
Chengzhu Yu, Heng Lu, Na~Hu, Meng Yu, Chao Weng, Kun Xu, Peng Liu, Deyi Tuo,
  Shiyin Kang, Guangzhi Lei, et~al.,
\newblock ``Durian: Duration informed attention network for speech
  synthesis.,''
\newblock in {\em INTERSPEECH}, 2020, pp. 2027--2031.

\bibitem{ren_fastspeech_2021}
Yi~Ren, Chenxu Hu, Xu~Tan, Tao Qin, Sheng Zhao, Zhou Zhao, and Tie-Yan Liu,
\newblock ``Fastspeech 2: Fast and high-quality end-to-end text to speech,''
\newblock {\em arXiv preprint arXiv:2006.04558}, 2020.

\bibitem{zhang_learning_2019}
Yu~Zhang, Ron~J Weiss, Heiga Zen, Yonghui Wu, Zhifeng Chen, RJ~Skerry-Ryan,
  Ye~Jia, Andrew Rosenberg, and Bhuvana Ramabhadran,
\newblock ``Learning to speak fluently in a foreign language: Multilingual
  speech synthesis and cross-language voice cloning,''
\newblock {\em arXiv preprint arXiv:1907.04448}, 2019.

\bibitem{he_multilingual_2021}
Mutian He, Jingzhou Yang, and Lei He,
\newblock ``Multilingual byte2speech text-to-speech models are few-shot spoken
  language learners,''
\newblock {\em arXiv preprint arXiv:2103.03541}, 2021.

\bibitem{li_bytes_2018}
Bo~Li, Yu~Zhang, Tara Sainath, Yonghui Wu, and William Chan,
\newblock ``Bytes are all you need: End-to-end multilingual speech recognition
  and synthesis with bytes,''
\newblock in {\em ICASSP 2019-2019 IEEE International Conference on Acoustics,
  Speech and Signal Processing (ICASSP)}. IEEE, 2019, pp. 5621--5625.

\bibitem{zhao_towards_2020}
Shengkui Zhao, Trung~Hieu Nguyen, Hao Wang, and Bin Ma,
\newblock ``{Towards Natural Bilingual and Code-Switched Speech Synthesis Based
  on Mix of Monolingual Recordings and Cross-Lingual Voice Conversion},''
\newblock in {\em Proc. Interspeech 2020}, 2020, pp. 2927--2931.

\bibitem{cao_code-switched_2020}
Yuewen Cao, Songxiang Liu, Xixin Wu, Shiyin Kang, Peng Liu, Zhiyong Wu, Xunying
  Liu, Dan Su, Dong Yu, and Helen Meng,
\newblock ``Code-switched speech synthesis using bilingual phonetic
  posteriorgram with only monolingual corpora,''
\newblock in {\em ICASSP 2020-2020 IEEE International Conference on Acoustics,
  Speech and Signal Processing (ICASSP)}. IEEE, 2020, pp. 7619--7623.

\bibitem{hemati_using_2020}
Hamed Hemati and Damian Borth,
\newblock ``Using ipa-based tacotron for data efficient cross-lingual speaker
  adaptation and pronunciation enhancement,''
\newblock {\em arXiv preprint arXiv:2011.06392}, 2020.

\bibitem{zhan_improve_2021}
Haoyue Zhan, Haitong Zhang, Wenjie Ou, and Yue Lin,
\newblock ``Improve cross-lingual text-to-speech synthesis on monolingual
  corpora with pitch contour information,''
\newblock in {\em Interspeech 2021}. pp. 1599--1603, {ISCA}.

\bibitem{whitehill_multi-reference_2020}
Matt Whitehill, Shuang Ma, Daniel {McDuff}, and Yale Song,
\newblock ``Multi-reference neural {TTS} stylization with adversarial cycle
  consistency,''
\newblock in {\em Interspeech 2020}. pp. 4442--4446, {ISCA}.

\bibitem{xin_cross-lingual_2021}
Detai Xin, Yuki Saito, Shinnosuke Takamichi, Tomoki Koriyama, and Hiroshi
  Saruwatari,
\newblock ``Cross-lingual speaker adaptation using domain adaptation and
  speaker consistency loss for text-to-speech synthesis,''
\newblock in {\em Interspeech 2021}. pp. 1614--1618, {ISCA}.

\bibitem{schroff_facenet_2015}
Florian Schroff, Dmitry Kalenichenko, and James Philbin,
\newblock ``Facenet: A unified embedding for face recognition and clustering,''
\newblock in {\em Proceedings of the IEEE conference on computer vision and
  pattern recognition}, 2015, pp. 815--823.

\bibitem{rallabandi_building_2017}
{SaiKrishna} Rallabandi and Alan~W. Black,
\newblock ``On building mixed lingual speech synthesis systems,''
\newblock in {\em Interspeech 2017}. pp. 52--56, {ISCA}.

\bibitem{csmsc}
D.~B. China,
\newblock ``Chinese standard mandarin speech corpus,''
  \url{https://www.data-baker.com/open source.html}, 2017.

\bibitem{LJSpeech17}
Keith Ito and Linda Johnson,
\newblock ``The lj speech dataset,''
  \url{https://keithito.com/LJ-Speech-Dataset/}, 2017.

\bibitem{povey_kaldi_nodate}
Daniel Povey, Arnab Ghoshal, Gilles Boulianne, Lukasˇ Burget, Ondˇrej
  Glembek, Nagendra Goel, Mirko Hannemann, Petr Motlıcˇek, Yanmin Qian, Petr
  Schwarz, Jan Silovsky, Georg Stemmer, and Karel Vesely,
\newblock ``The kaldi speech recognition toolkit,''
\newblock p.~4.

\bibitem{kong_hifi-gan_2020}
Jungil Kong, Jaehyeon Kim, and Jaekyoung Bae,
\newblock ``Hifi-gan: Generative adversarial networks for efficient and high
  fidelity speech synthesis,''
\newblock in {\em Advances in Neural Information Processing Systems},
  H.~Larochelle, M.~Ranzato, R.~Hadsell, M.~F. Balcan, and H.~Lin, Eds. 2020,
  vol.~33, pp. 17022--17033, Curran Associates, Inc.

\end{thebibliography}

\end{document}